\documentclass[pra,superscriptaddress,showpacs,12pt]{revtex4}
\usepackage{graphicx}
\usepackage{amssymb}

\begin{document}

\title{Operator representations for a class of quantum entanglement measures and criterions}

\author{Z. Xu}
\affiliation{Department of Physics, Tsinghua University, Beijing,
100084, China} \affiliation{Center for Advanced Study, Tsinghua
University, Beijing 100084, China}

\author{B. Zeng}
\affiliation{Department of Physics, Tsinghua University, Beijing,
100084, China}

\author{D.L. Zhou}
\affiliation{Center for Advanced Study, Tsinghua University,
Beijing 100084, China}

\date{\today }

\begin{abstract}
We find that a class of entanglement measures for bipartite pure
state can be expressed by the average values of quantum operators,
which are related to any complete basis of one partite operator
space. Two specific examples are given based on two different ways
to generalize Pauli matrices to $d$ dimensional Hilbert space and
the case for identical particle system is also considered. In
addition, applying our measure to mixed state case will give a
sufficient condition for entanglement.
\end{abstract}


\maketitle

\section{Introduction}
Quantum entanglement is an essential physical resource to process
quantum information and computation, which enables us to complete
the tasks intractable in classical domain, such as quantum
teleportation, quantum cryptography, Shor's algorithm of factoring
large numbers, and Grover's quantum searching algorithm \cite{Ni}.

In order to use such kind of resources efficiently, it is
necessary to qualify the properties and quantify the degrees of
quantum entanglement for a given quantum state. In this direction,
continuous progresses have been made. To clarify the meaning and
qualify the properties of entanglement, Werner defined separate
state from whether being able to prepare the state classically,
whose definition has become the standard mathematical basis of
entanglement state\cite{Wer}. Next, Peres proposed a famous
necessary condition for separity
---positive of partial transpose density operator \cite{Per}, then
Horodeckies prove this criteria is also sufficient in the cases of
$\mathcal{H}^2\otimes \mathcal{H}^2$ and $\mathcal{H}^2\otimes
\mathcal{H}^3$ \cite{Ho1}.

In order to quantify this property, many entanglement measures
have been proposed in the past years, both for pure states and
mixed states\cite{Be,Wo,Ho}. However, only for bipartite pure
states the quantitative theory of entanglement satisfies with all
the \textit{priori} axioms of a good entanglement measure, which
is mainly due to the existence of the celebrated Schmidt
decomposition for these states. It is well known that the von
Noemann entropy of the reduced density matrix $S_{E}$ is the
unique measure for bipartite pure states in the sense that $S_{E}$
can be concentrated and diluted with unit asymptotic
efficiency\cite{Be,Ni1}. However, Vidal developed the concept of
entanglement monotone and shows that to characterize the non-local
property of finite number bipartite pure states, indeed $d-1$
independent measure is needed in the sense that there is $d-1$
Schmidt coefficients \cite{Vi}. Thus it is known that although the
entanglement monotones have different asymptotic properties than
$S_{E}$, they are important for characterizing non-local
properties under LOCC transformations. For mixed state case,
recently Wooters gives out an explicit expression of entanglement
of formation in $\mathcal{H}^2 \otimes \mathcal{H}^2$\cite{Woo}.
However, there are still many open questions, especially for many
partite system and mixed states.

As we know, entanglement measures are functionals of density
operator. However, quantities in traditional quantum physics are
quantum observables. In this sense, entanglement measures are not
standard physical quantum observables, and they are also not the
average values of some entanglement measure operators. In this
article, we attempt to establish the relations between
entanglement measures and quantum observables. However, we achieve
this end only for a specific class of entanglement measures, which
will be analyzed in Section $2$, where the case for identical
system is also considered. In sec. $3$, two specific examples are
given based on different generations of Pauli matrixes, which
preserve Hermitian and Unitary respectively. Finally, we apply
those results to form a criterion of mixed state entanglement and
a short summary is also given in sec. $4$.

\section{Operator space representations for a class of quantum entanglement measures}

For a bipartite pure state $|\psi_{AB}\rangle$ in Hilbert space
$\mathcal{H}_A^d \otimes \mathcal{H}_B^{d^\prime} \quad (d\le
d^{\prime})$, a class of functions of reduced density operator
$\rho_A$ can be defined as
\begin{equation}
M_e(n)=1-\mbox{Tr} {\hat{\rho}_A}^n, \qquad (n\in N\ and\ n\ge 2)
\end{equation}
where the reduced density operator $\hat{\rho}_A=\mbox{Tr}_B
(|\psi_{AB}\rangle \langle \psi_{AB}|)$. It is easy to show that
the above class of functions are entanglement monotones, or
entanglement measures due to the fact that they only depend on the
eigenvalues the reduced density matrix $\rho_A$, or equivalently
the Schmidt numbers of the state $|\psi_{AB}\rangle$ \cite{Vi}.

Denote the linear space of operators act on the Hilbert space
$\mathcal{H}_A^d$ as $\mathcal{M}_A^d$, which is a linear space of
$d\times d$ dimensions, and denote the arbitrary operator $P\in
\mathcal{M}_A^d$ as $|P\rangle$ and $P^{\dag}\in \mathcal{M}_A^d$
as $\langle P|$. Define the inner product of $\mathcal{H}_A^d$ as
\begin{equation}
\langle P | Q \rangle =\mbox{Tr} (P^\dagger Q), \qquad \forall P,Q
\in \mathcal{M}_A^d.
\end{equation}
Then we can rewrite the class of entanglement measures in Eq. (1)
as
\begin{equation}
M_e(n)=1-\langle \rho_A| \rho_A^{n-2} | \rho_A \rangle.
\end{equation}

For each entanglement measure $M_e(n)$, we take $n-1$ sets of
complete operators $ S_C^m=\{ \mathcal{O}_i^m \} \quad
(m=1,2,\cdots,n-1) $, which satisfy
\begin{equation}
\sum_i |\mathcal{O}^m_i\rangle \langle \mathcal{O}^m_i|=1.
\end{equation}
Using the above relations, we rewrite the entanglement measures as
\begin{eqnarray}
M_e(n)&=&1- \sum_{i_1,i_2,\cdots,i_{n-1}}\langle
\rho_A|\mathcal{O}^1_{i_1} \rangle \langle\mathcal{O}^1_{i_1}
|\rho_A |\mathcal{O}^2_{i_2}\rangle \cdots
\langle\mathcal{O}^{n-2}_{i_{n-2}}
|\rho_A|\mathcal{O}^{n-1}_{i_{n-1}}\rangle
\langle\mathcal{O}^{n-1}_{i_{n-1}}|\rho_A \rangle \nonumber\\
&=&1- \sum_{i_1,i_2,\cdots,i_{n-1}}\langle \mathcal{O}^1_{i_1}
\rangle \langle\mathcal{O}^2_{i_2}{\mathcal{O}^1_{i_1}}^\dagger
\rangle \cdots
\langle\mathcal{O}^{n-1}_{i_{n-1}}{\mathcal{O}^{n-2}_{i_{n-2}}}^\dagger
\rangle \langle{\mathcal{O}^{n-1}_{i_{n-1}}}^\dagger\rangle,
\label{mr}
\end{eqnarray}
where
\begin{equation}
\langle \mathcal{O} \rangle=\mbox{Tr} (\rho_A \mathcal{O}),
\end{equation}
and obviously it is the also expected value of operator
$\mathcal{O}$ in state $|\psi_{AB}\rangle$.

Eq. (\ref{mr}) is the main result of this paper and it relates the
entanglement measures with physical obeservables, \textit{i.e.},
it tells us the following information: If we obtain a serious of
expected values for some complete operators, the degree of
entanglement can be evaluated by Eq. (\ref{mr}). In other words,
we can measure entanglement by measuring some physical
observables. It is worthy to note that physical observables can be
represented by unitary operators besides Hermitian ones, in the
sense that for any unitary operator we can always find such a
Hermitian operator that might be mapped to the unitary operator by
exponential functions.

In the case of $n=2$, Eq.(\ref{mr}) takes a much simpler form:
\begin{equation}
M_e(2)=1-\sum_i {\left| \langle \mathcal{O}_i \rangle
\right|}^2=\frac{1}{2}C_I^2.
\end{equation}

Where $C_I$ is the generalized concurrence, or $I$-concurence for
two qudits. It is well known that among all the entanglement
monotones, concurrence is important since it is related to the
entanglement of formation for two qubits\cite{Wo}. It is also
found that there is many ways to define concurrence for bipartite
pure states, which reveals different physical
meanings\cite{Be,Ab,Ch}. Very recently, the concept of concurrence
is generalized to higher dimensions based on the ``universal
inverter" and the mathematical point of view\cite{Ru,Fei},
although almost all the ways of defining concurrence for two
qubits can not be generalized to higher dimensions\cite{We}. It is
found that the generalized $I$-concurence $C_I$ with its mixed
state counterpart is useful in characterizing the non-local
properties for bipartite states, both pure and mixed\cite{Ru2,De}.
Due to these reasons, we will concentrate our attention to this
specific case and give explicit examples in the following section.

Before going to concrete example, we first consider a special
case, \textit{i.e.}, entanglement of identical particle systems.
Although the theory of entanglement is widely developed in the
systems of distinguishable particles, only very recently the
entanglement properties in identical particle systems began to
attract much attention\cite{slm,sckll,You,lbll,gf,esbl} in the
fields of quantum information and quantum computation. It is also
shown that for any $N$ identical particle pure state
$|\psi_N\rangle$, all the information of their quantum correlation
between one particle and the others are contained in the single
particle density matrix \cite{fan}. Therefore our entanglement
measure is not only suitable for bipartite case here, but also a
measure (to see this is indeed an entanglement measure here, see
ref \cite{Bre}) for $N$ identical particle entanglement,
\textit{i.e.},

\begin{equation}
M_e(2)=1-\sum\limits_{i=0}^{d^2-1}|\langle\Psi_N|O_{i}|\Psi_N\rangle|^2.
\end{equation}

\section{Realization of $M_e(2)$ with Pauli Operators and its high dimensional generalizations}

In this section, we will give examples of realization of $M_e(2)$
with Pauli operators and the two different generations of Pauli
matrixes to higher dimensional Hilbert space, which preserve
Hermitian and Unitary respectively. We know that an arbitrary
state of two qubits in the Hilbert space $H=H_{A}{\otimes}H_{B}$
(where $H_{A}=H_{B}=C^{2}$) can be written as

\begin{equation}
\Psi=\alpha_{1}|00\rangle+\alpha_{2}|01\rangle+\alpha_{3}|10\rangle+\alpha_{4}|11\rangle.
\end{equation}
where $\sum_{i}|\alpha_{i}|^{2}=1$.

Let $s_i=\frac{1}{\sqrt{2}}\sigma_i$, where $\sigma_0=I$ and
$\sigma_{i}\ (i=1,2,3)$ are usual Pauli operators. Obviously
$\{s_i\}$ form a basis for $2\times 2$ operator and thus

\begin{eqnarray}
M_e(2)&=& 1-\sum\limits_{i=0}^{3} \langle s_i \rangle^2=
1-\frac{1}{2}\sum\limits_{i=0}^{3}
\langle \sigma_i \rangle^2\nonumber\\
&=&\frac{1}{2}\left( 1-\sum\limits_{i=0}^{3} \langle \sigma_i
\rangle^2 \right)=\frac{1}{2}C^2,
\end{eqnarray}
where

\begin{equation}
C=2|\alpha_{1}\alpha_{4}-\alpha_{2}\alpha_{3}|
\end{equation}
is the usual concurrence.

For qudits case, we demonstrate two kinds of commonly used
``generalized" Pauli operators. The first kind is so-called
Gell-mann matrices $\lambda_i$, which are Hermitian generators of
$SU(d)$. From the completeness relation of $\lambda_i$

\begin{equation}
\sum\limits_{i=1}^{d^2-1}(\lambda_i)_{kl}(\lambda_i)_{pq}=2\left(\delta_{kp}\delta_{lp}-
\frac{1}{d}\delta_{kp}\delta_{lp}\right),
\end{equation}
it is easy to show that

\begin{equation}
M_e(2)=\frac{(d-1)}{d}-\frac{1}{d}\sum\limits_{i=1}^{d^2-1}
\langle\Psi|\lambda_{i}|\Psi\rangle^2.
\end{equation}
It is noticed that this result is in fact already gotten in Ref.
\cite{Mah}.

Another kind of generalized Pauli operators are $Z^mX^n$, which
are all unitary matrices. Here $Z$ and $X$ are the generators of
quantum plane algebra with $q^{d}=1$ \cite{Sun}. The $Z$-diagonal
representation of $Z$ and $X$ given by
\begin{eqnarray}
Z &\equiv &\sum_{k_{0}}^{d-1}|k\rangle q_{d}^{k}\langle k|, \\
X &\equiv &\sum_{k=0}^{d-1}|k\rangle \langle k+1|,
\end{eqnarray}%
for $q_{d}=e^{i\frac{2\pi }{d}}$.

From the completeness relation of $Z^mX^n$

\begin{equation}
\frac {1} {d} \sum\limits_{m,n=0}^{d-1}|Z^mX^n\rangle\langle
Z^mX^n|=1,
\end{equation}
it is easy to show that

\begin{equation}
M_e(2)=1-\frac {1} {d}
\sum\limits_{m,n=0}^{d-1}|\langle\Psi|Z^mX^n|\Psi\rangle|^2.
\end{equation}

\section{Applications to mixed state entanglement}

Apparently the entanglement measure defined in Eq. (1) cannot be a
entanglement measure for mixed state case. However, the technique
developed above can help us to derive some criterion for mixed
state entanglement.

The completeness relation Eq. (4) is equivalent to

\begin{equation}
\sum\limits_{i=1}^{d^2}O_{i}^{\dag}YO_i=trY
\end{equation}
for arbitrary $d\times d$ operator $Y$. Therefore, if $Y=I$ and
$O_m^i$ are hermitian, we have

\begin{equation}
\sum\limits_{i=1}^{d^2}O_i^2=d.
\end{equation}
So the sum of uncertainty of $O_m$ gives that

\begin{eqnarray}
\sum\limits_{i=1}^{d^2}(\delta O_i)^2
&=&\sum\limits_{i=1}^{d^2}\mbox{tr}(\rho
O_i^{2})-(\mbox{tr}(\rho O_{i}))^2\nonumber\\
&=&d-\sum\limits_{i=1}^{d^2}(\mbox{tr}(\rho O_{i}))^2=d-\sum\limits_{i=1}^{d^2}\langle O_{i}\rangle^2\nonumber\\
&=&d-\mbox{tr}(\rho^2)\geq d-1.
\end{eqnarray}

Then we can get a non-trivial sum uncertainty relation \cite{Hof1}

\begin{equation}
\sum\limits_{i=1}^{d^2}\Big(\delta\left(O_{iA}-O_{iB}\right)\Big)^2\geq
2(d-1)
\end{equation}
to result in a sufficient condition for entanglement if the above
inequality is violated. This entanglement criterion may be
stronger than Peres-Horodecki criterion for it is shown that some
PPT state violate this criterion\cite{Hof2}.

This idea is also useful in $N$-identical particle case, which
will lead to an entanglement criterion based on the sum
uncertainty of collective operators for many identical particles.
For $N$ identical particles, the collective operator is defined as

\begin{equation}
O_i=\sum\limits_{K=1}^{N}O_{iK},\; (K=1,2,...,N).
\end{equation}
Correspondingly the sufficient condition for a $N$ identical
particles state to be entangled is

\begin{equation}
\sum\limits_{i=1}^{d^2}(\delta O_{i})^2< N(d-1).
\end{equation}

Usually, for $N$-identical qubits we choose $O_{iK}\, (i=0, 1, 2,
3)$ as $I,s_1,s_2,s_3$, then $O_i\, (i=0, 1, 2, 3)$ will be the
total spin of the system apart from a constant multiplier
$\frac{1}{\sqrt{2}}$. This criterion is analogous to the
criterions defined by the squeezing parameters in the literatures
\cite{sq1,sq2,sq3}.

{\vskip 2mm}

In summary, we showed that a class of entanglement measures for
bipartite pure state can be expressed by the average values of
quantum operators, which are related to any complete basis of one
partite operator space with two specific examples given based on
two different ways to generalize Pauli matrices to $d$ dimensional
Hilbert space. In addition, applying our measure to mixed state
case gave a sufficient condition for entanglement and the case for
identical particle systems was also considered.

\begin{acknowledgements}

The authors would like to thank Prof. L. You for useful
discussions. The work of Z. X is supported by CNSF (Grant No.
90103004, 10247002). The work of D. L. Z is partially supported by
the National Science Foundation of China (CNSF) grant No.
10205022.

\end{acknowledgements}

\end{document}